# SPECTRUM SENSING STRATEGY TO ENHANCE THE QOS IN WHITE-FI NETWORKS

Nabil Giweli, Seyed Shahrestani and Hon Cheung

School of Computing, Engineering and Mathematics, Western Sydney University, Sydney, Australia

## ABSTRACT

*The rapidly growing number of wireless devices running applications that require high bandwidths, has resulted in increasing demands for the unlicensed frequency spectrum. Given the scarcity of allocated unlicensed frequencies, meeting such demands can become a serious concern. Cognitive Radio (CR) technology opens the door for the opportunistic use of the licensed spectrum to partially address the issues relevant to the limited availability of unlicensed frequencies. Combining CR and Wi-Fi to form the so-called White-Fi networks, has been proposed for achieving higher spectrum utilization. This article discusses the spectrum sensing in White-Fi networks and the impacts that it has on the QoS of typical applications. It also reports the analysis of such impacts through various simulation studies. Our results demonstrate the advantages of an adaptive sensing strategy that is capable of changing the related parameters based on QoS requirements. We also propose such a sensing strategy that can adapt to the IEEE 802.11e requirements. The goal of the proposed strategy is the enhancement of the overall QoS of the applications while maintaining efficient sensing of the spectrum. Simulation results of the scenarios that implement the proposed mechanisms demonstrate noticeable QoS improvements compared to cases where common sensing methods are utilized in IEEE802.11 networks.*

## 1. INTRODUCTION

Traditionally, the radio spectrum is statically divided into frequency bands, most of which are licensed to organizations and companies usually referred to as primary users (PUs). A small number of frequency bands are unlicensed including the unlicensed Industrial, Scientific, and Medical (ISM) bands, which are used by a variety of indoor and short-range wireless communication systems, such as Wi-Fi, Bluetooth, and Zigbee. These free unlicensed frequencies are not sufficient to handle the rapidly growing number of wireless devices using these unlicensed bands. Also, modern applications running on these devices demand more bandwidth. These modern applications usually involve multimedia communications, e.g., media streaming, video conferencing, and interactive gaming. One of the promising solutions to increase radio frequency spectrum availability to these wireless devices is to add the Cognitive Radio (CR) capability to such devices. With CR capability, a wireless device can operate opportunistically, as a Secondary User (SU), over licensed frequency channels when they are unused, i.e., White Spaces (WS) or spectrum holes in the licensed bands. Television (TV) bands are the most attractive frequency ranges for such opportunistic use of the spectrum by SUs. This is because the TV bands show high availability of WSs and their scheduled use by the PUs can be obtained through Geo location Database (GDB) services[1]. Moreover, the TV spectrum is located below 1 GHz. Compared to the higher ISM bands; these frequencies offer more desirable propagation characteristics. An IEEE-802.11 protocol with CR capability is often referred to as CR Wi-Fi, White-Fi, Wi-Fi Like or IEEE-802.11af. The IEEE 802.11af is the first draft standard for CR networks based on IEEE 802.11 to operate in TV WS[2]. Based on the IEEE 802.11af standard, the White-Fi devices can communicate through either ISM channels or TV WSs. Figure 1shows the main approaches for assessing the potential operation bands along with their related basic conditions and required actions.

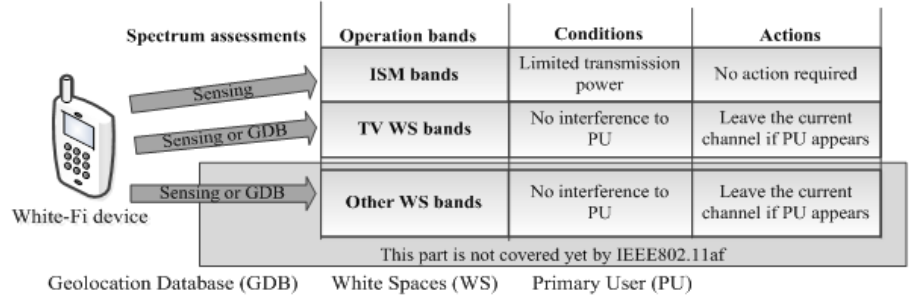

Figure1.White-Fi potential operation bands and their required conditions and actions

Primarily, the CR capabilities and requirements are established at the Physical (PHY) and Medium Access Control (MAC) layers of wireless systems. Spectrum sensing is one of the most important functions in CR to identify the available spectrum holes and to protect the PU from interference. Conducting sensing has its impact on transmission delays and throughput. The Carrier Sense Multiple Access with Collision Avoidance (CSMA/CA) mechanism is used in wireless devices based on various 802.11 standards to share the ISM bands[3]. The concept is that a Wi-Fi device checks the channel occupancy before transmitting over it. Typically, this checking is accomplished by using a simple sensing technique, e.g., Energy Detection (ED), where the energy of the channel is measured and compared to a predefined threshold. If the measured energy level exceeds the threshold, the channel is in use by another device; otherwise, it is idle.

The remaining parts of this article are organized as follows. In section 2, some of the previous related works are reviewed and compared to the contributions of this study. Correlation between QoS requirements and sensing is discussed in Section 3.Then the impacts of sensing duration on QoS are identified and analyzed by simulations in Section 4. In Section 5,the proposed solution for enhancing QoS in White-Fi networks by selecting sensing strategy based on QoS requirements is demonstrated. The proposed sensing strategy is evaluated in Section 6. Conclusions and future works are described in Section 7.

## 2. RELATED WORK AND MOTIVATIONS

The success of White-Fi technology highly depends on the QoS level that it can offer for various communication applications. Although the QoS in CR networks, in general, has drawn more attention recently, White-Fi networks have received a minor portion of that attention. Articles that have been published to study QoS issues in White-Fi networks based on sensing are related to our work in this paper. The most related work is that where the sensing duration and its effect on QoS are considered.

The effect of the sensing duration on the access delay of an SU was analyzed when the Request-to-Send/Clear-to-Send(RTS/CTS)mechanism was enabled for spectrum access in[4]. The optimal sensing length was formulated as a function of the false alarm and miss detection probabilities and the number of SUs contending for the same access point. The ED is the main spectrum assessment method, and the maximum sensing duration used for analysis was 25ms.The sensing was only conducted at the beginning of the contention period and before sending an RTS packet. The study found that the access delay of an SU depends on the sensing length, and its optimal value varies based on the number of contending SUs.

The relation and the possible trade-off between the spectrum sensing duration and achievable throughput of SUs were addressed in several studies [5, 6]. The approximated analytical formulation of the optimally saturated throughputs for multiple SUs based on CSMA/CA mechanism was proposed in[14]. The discrete-time Markov chain was used to model this formulation for different false alarm probabilities and a varying number of SUs. The numerical results of that work showed the significant role of sensing in improving the throughput achieved by the SUs. They have also demonstrated how the saturated throughput was affected by the sensing false alarm probability. The analysis was approximate, using typical parameters. The main aim of that work was to conduct an initial study of the performance of CR techniques employed in 802.11-based networks. A similar study, aiming to find the optimal sensing duration that can achieve the maximum throughput under unsaturated traffic conditions, has also been conducted[6].In this work, the discrete-time Markov chain model was used by the authors to model their proposed MAC structure for analyzing the performance of SUs. The sensing was conducted by SUs only at the beginning of the contention period when one access point and several SUs exist. The optimal sensing duration was investigated within the range of 0.5 ms to 3 ms for a different number of SUs (5, 10, 15 and 20) of various queuing probabilities and contention window sizes. Other conditions, such as the false alarm probability, detection probability and single-to-noise ratio (SNR), were assumed constant and the same for all SUs. Based on the assumptions above, the optimal sensing duration was found to be around 2 ms for all the simulated scenarios under certain assumptions and conditions[6]. Therefore, an optimal sensing duration can be determined for maximum throughput under unsaturated traffic conditions.

The studies mentioned above have been mainly based on the use of ED as their spectrum assessment method, with relatively small sensing periods. Practically, the ED method performs poorly in low SNR environments, and cannot distinguish between PU signals and other SU signals. The use of higher accuracy spectrum assessment methods implies the need of longer sensing durations. In our study, a wider range of sensing durations is considered to reflect the potential use of more complex spectrum sensing methods.

## 3. CORRELATIONS BETWEEN QOS AND SENSING

The IEEE 802.11e standard is proposed to enhance QoS in IEEE 802.11 networks [7]. As applications have different requirements, in 802.11e, the frames belonging to different applications are prioritized with one of the eight user priority (UP) levels. In contrast, previous IEEE 802.11 standards use the Distributed Coordination Function (DCF) mechanism at MAC layer where the best-effort service is provided equally to all traffic streams from different applications to access the medium. In IEEE 802.11e, the Hybrid Coordination Function (HCF) is used for prioritizing traffic streams to enhance QoS on top of the DCF. The HCF accommodates two medium access methods, i.e., a distributed contention-based channel access mechanism, called Enhanced Distributed Channel Access (EDCA), and a centralized polling-based channel access mechanism, called HCF Controlled Channel Access (HCCA). Based on the UP, the EDCA defines four Access Categories (AC); voice (AC_VO), video (AC_VI), best-efforts (AC_BE) and background (AC_BK). These categories are assigned different priorities ranging from highest to lowest respectively. The category AC_VO has top priority and is usually given to traffic carrying voice information. It is followed by the AC_VI category for video traffic and then the AC_BE category for data traffic. The category AC_BK has the lowest priority and is usually assigned to unnecessary data traffic.

Each AC has a Contention Window (CW) that has a specified minimum size and maximum size, i.e., $CW_{min}[AC]$ and $CW_{max}[AC]$.The values $CW_{min}[AC]$ and $CW_{max}[AC]$ are calculated for each AC based on the predefined values $PHY\_CW_{min}$ and $PHY\_CW_{max}$ for the variousPHY layers supported by the IEEE 802.11e. Also, an Arbitration InterFrame Space (AIFS) value and a Transmit Opportunity (TXOP) interval are used to support QoS prioritization [8]. Instead of using fixed Distributed Inter-Frame Space (DIFS), also called DCF inter-frame space, the AIFS[AC] value is a variable value calculated based on the AC. For instance, possible values for AIFS[AC] are; AIFS[AC_BK], AIFS[AC_BE], AIFS[AC_VI] or AIFS[AC_VO]. The AIFS value determines the time that a node defers access to the channel after a busy period, and before starting or resuming the back-off duration. Hence, the time for a station to wait for the channel to become idle before it starts sending data is calculated based on the AC category of the data [9]. However, the Short InterFrame Space (SIFS) value is used as the shortest InterFrame Space (IFS) value for transmitting high priority frames, such as those for acknowledgments. Therefore, the higher priority frames access the operational channel earlier than other frames in the same transmission queue. For the frames belonging to the categories of AC_VI and AC_VO, the AIFS, and CW values are set smaller than those for the frames of the categories AC_BE and AC_BK. This mechanism reduces the delays of frames carrying multimedia information and improving the QoS of the underlying applications.

White-Fi users and SUs based on other different wireless technologies may coexist in the same available WSs. Consequently, a White-Fi user needs to distinguish between three types of users, i.e., the PU, the other White-Fi users and the other non-White-Fi SUs. The simple sensing methods, e.g., ED, cannot distinguish amongst the signals related to these users. Advanced techniques, such as Matched Filter Sensing (MFS) method, can distinguish between signals when prior information about these signals is available .However, for proper operation, they require extended sensing durations. In the case of White-Fi, increasing sensing time will impact the effectiveness of the IEEE802.11e standard on improving QoS in IEEE802.11 networks. The complications of the sensing operation should be investigated under different settings of the associated parameters. In the following section, the reported simulation results demonstrate how increasing the sensing duration affects the effectiveness of the IEEE 802.11e mechanisms in White-Fi networks.

## 4. SIMULATION STUDIES: IMPACTS OF SENSING DURATION

To explore the effects of sensing function on the frame transmission delays that impact the QoS of applications running on a White-Fi device, a simulation tool Modeler 18.0 from Riverbed (formerly Opnet) [10] is used. Modeler 18.0 supports different types of applications and network traffics. However, CR networks are not implemented in the Modeler 18.0. Hence, we have customized the standard Wi-Fi node to include a function to work with different sensing periods to simulate the behavior of a White-Fi node. The settings of some common parameters of all simulation scenarios are shown in Table 1. The IEEE 802.11e was supported in all scenarios with the settings illustrated in Table 1. Simulations were conducted for different sensing durations and different application categories. The main three traffic types considered were the voice, video conferencing, and email. For a voice application, voice traffic was generated as IPv4 unicast traffic flows between the nodes. For generating email and video conferencing traffic, a server was used to run these applications in the infrastructure mode network scenarios.

International Journal of Computer Networks & Communications (IJCNC) Vol.10, No.1, January 2018

Table 1 Simulation parameters settings

| Parameter | value |
|---|---|
| Data rate | 26 Mbps / 240Mbps |
| Buffer Size | 256000 bits |
| Maximum transmitter frame size | 3839 bytes |
| Maximum acceptable frame size | 8191 bytes |
| **EDCA Parameters:** | |
| Voice: | $CW_{min} = (PHYCW_{min} + 1) / 4 - 1$ |
| | $CW_{max} = (PHYCW_{min} + 1) / 2 - 1$ |
| | AIFSN = 2 |
| Video: | $CW_{min} = (PHYCW_{min} + 1) / 2 - 1$ |
| | $CW_{max} = PHYCW_{min}$ |
| | AIFSN = 2 |
| Best Effort: | $CW_{min} = PHYCW_{min}$ |
| | $CW_{max} = PHYCW_{max}$ |
| | AIFSIN = 3 |
| Background: | $CW_{min} = PHYCW_{min}$ |
| | $CW_{max} = PHYCW_{max}$ |
| | AIFSIN = 7 |

The resulted values were captured under bucket mode with a sample mean of 100 values per a result statistic. In the Bucket mode, the data is collected at all of the points over the time interval or sample count into a "data bucket" and generates a result from each bucket. The wireless delay, simply called delay in this article, represents the end-to-end delay of all the data packets that are successfully received by the MAC layer and forwarded to the higher layer in a node. The media access delay is the sum of delays, including queuing and contention delays of all frames transmitted via the MAC layer. For each frame, the media access delay is calculated as the duration between the time when the frame is placed in the transmission queue until the time when the frame is sent to the physical layer for the first time. On other words, the media access delay is the time of processing a packet at the MAC layer.

### 4.1 SENSING DURATION AND QOS

In this subsection, the effect of sensing duration is analyzed for various applications. Three scenarios are implemented for that purpose: voice, email, and video. The sensing is conducted for all frames except response frames in all scenarios. For voice scenario, a four node Ad Hoc network is used, and IPv4 unicast voice traffic is generated amongst all the four nodes. Several simulations were conducted for different sensing durations from 1 ms to 300 ms. Also, the simulation was run when the sensing was neglected, i.e., 0 ms. As indicated in Figure 2,in the first few hundreds of seconds of the simulation time, the average delay sharply increases, irrespective of the employed sensing duration. The delaysthen begin to plateau for all cases.

International Journal of Computer Networks & Communications (IJCNC) Vol.10, No.1, January 2018

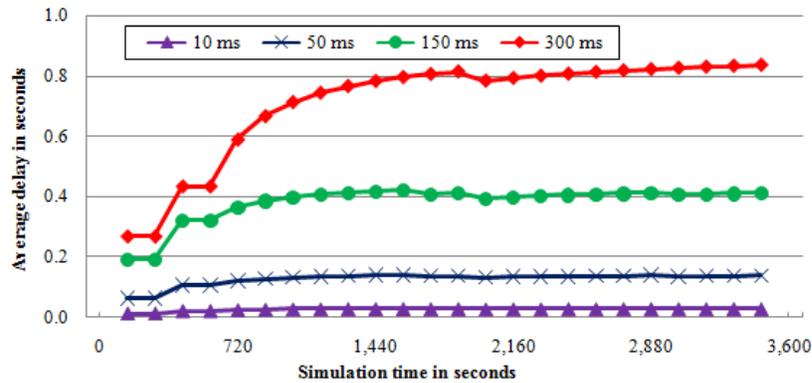

Figure 2. Average delay for different sensing durations (Ad Hoc network of four nodes with voice traffic)

In the email scenario, a server was added for simulating heavy email traffic between the server and other four nodes. Instead of operating as an ad hoc network in the previous scenario, the network, in this case,works in an infrastructure mode with the server also acting as an AP. Several simulations were run for different sensing periods from 1 ms to 300 ms. The simulated operation time of the network was an hour for each of these sensing durations. The simulation results are shown in Figure 3. According to the results, the maximum measured average delay is 0.35 seconds when the sensing length is 300 ms. The average delay is between 0.3 and 0.1 seconds for sensing durations between 250 ms and 100 ms while it is less than 0.1 seconds for sensing lengths less than 50 ms. These results show that the delay is less with heavy email traffic than the voice traffic.

The video application scenario was similar to the email application one, except that the added server was used for providing video conferencing application to the other four nodes. The server was used to generate high-resolution video conference traffic between the four nodes through the server. The simulations were conducted for different sensing durations from 1 ms to 300 ms. The average delay in this scenario has not exceeded 0.1 seconds even when the sensing length is 300 ms as shown in Figure 4.

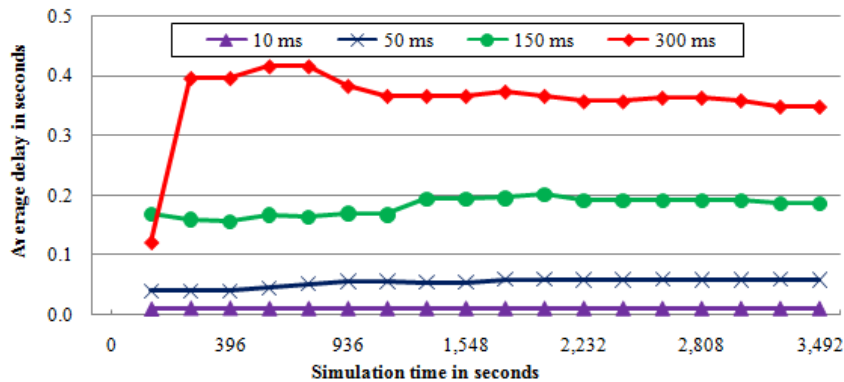

Figure3.Average delay for different sensing durations (Infrastructure network with Heavy email traffic)

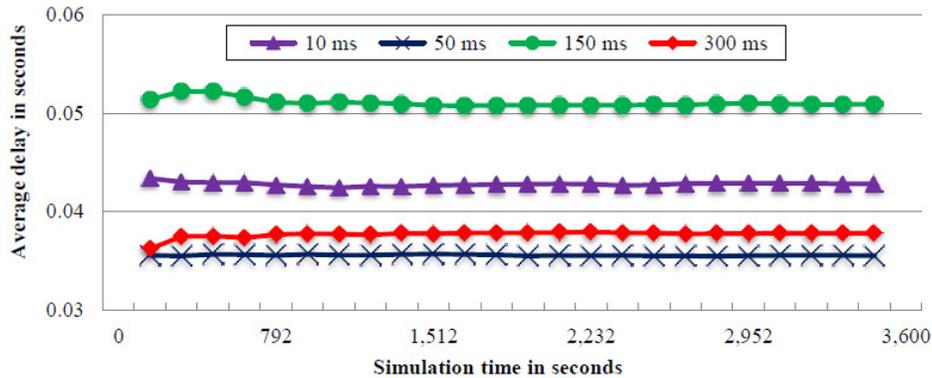

Figure 4. Average delay for different sensing durations (Infrastructure network with video conferencing traffic)

It can be seen that the 150 ms sensing duration caused the highest average delay, which was over 0.05 second and was even greater than the average delay for the cases of 300 ms sensing duration. The average delay in this scenario does not follow proportional relation between the average delay and sensing duration. An optimal sensing duration could be achieved for the video traffic under given conditions. For instance, Figure 4 shows that 50 ms and 300 ms sensing durations may lead to potential optimal values of the average delay.

## 4.2 OBSERVATIONS AND COMPARISONS

The comparison between the simulation results of the three traffic types, shown in Figure , illustrates that the voice traffic experiences the highest average delay. Compared to the other traffic, the voice is more sensitive to the length of the sensing period. The average delay in email traffic is proportional to the sensing duration, except with a smaller slope. Thus, the email traffic is less affected by the sensing duration than voice traffic. Moreover, email applications are not sensitive to delay. The video traffic is less susceptible to sensing length compared to the other traffics. In contrast, the average delay, and the sensing duration is not in a constant proportionality relation. The results in Figure demonstrate that the same sensing strategy has different effects on the traffics from various applications. As Figure shows, the sensing durations less than 50 ms cause an average delay of fewer than 0.2 seconds, which is acceptable in several applications. However, a longer sensing time could be required for achieving a higher accuracy. Although IEEE802.11e was enabled in the simulations, the results show that the voice and video traffic gain no benefit from it because of the impact of extended sensing time. Therefore, sensing duration and frequency should be conducted in a way that preserves the aim of improving QoS of different applications by using IEEE802.11e. In the next section, our proposed solution for enhancing QoS is discussed and evaluated by simulations.

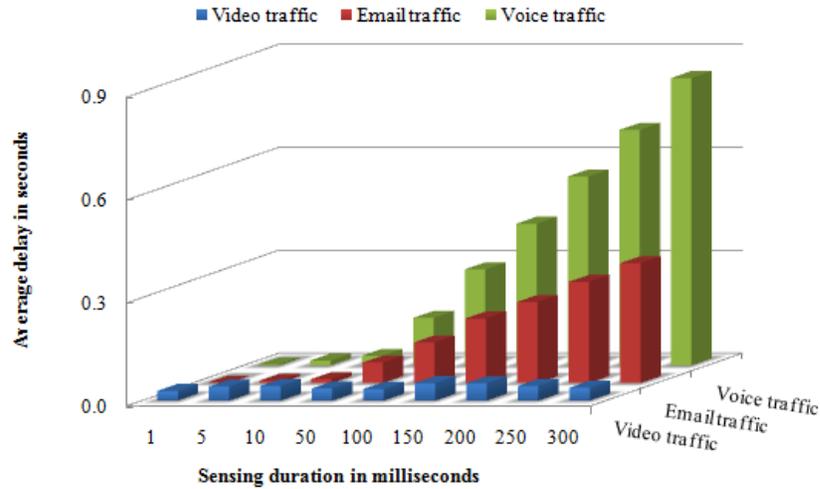

Figure 5.Comparison of the average delays experienced by various applications with varying sensing durations

## 5. SENSING STRATEGY AND ENHANCED QOS

The ED method is commonly used for sensing on CSMA/CA based networks because of its simplicity and low overhead. However, in a CR environment, the main ED drawback is its inability to recognize PU signal among other signals, and this reduces its sensing effectiveness. The MFS method can overcome this drawback and provide higher detection accuracy at the cost of more sensing time, complexity and power consumption. Therefore, the spectrum assessment method should be selected during operation by trading off between the sensing requirements and its implications. Such an approach imposes that the White-Fi device supports a set of different sensing techniques where each one of these methods is suitable for a particular set of operation requirements. Towards enhancing QoS in White-Fi networks, a mechanism of selecting the proper sensing method and MAC operation settings that suit the QoS application requirements is proposed.

The required QoS for different applications is classified into four levels based on the four ACs in IEEE 802.11e as shown in Table . The spectrum assessment operation is categorized into four types; Coarse (C), Moderate (M), Fine (F) and Extra Fine (EF) sensing. Each sensing type will be conducted based on the AC of the frame to be transmitted as shown in Table . The proposed Sensing duration $S_d$ range for each type is chosen based on our previous study and classification of different spectrum assessment methods [11]. In the C sensing type, the duration $S_d$ is less or equal 1 ms to give higher priority to AC_VO frames with less impact on the delay. However, the spectrum assessment method that can be used within this short time is the blind sensing method, such as ED. Consequently, the C sensing type cannot distinguish between PU and other SU signals, and poor accuracy is expected at low SNR. In the M sensing type, the spectrum assessment methods that can be used for $S_d$ values between 1 and 5 ms are similar to those for the C category. The use of the M type results in a slight improvement in sensing accuracy, particularly at low SNR. Therefore, AC_VI frames have less priority than AC_VO to win the contention window, and more delay is predicted. For the F sensing type, the duration $S_d$ is larger than 5 ms up to 50 ms. Thus, spectrum assessment methods that can differentiate PU singles from other signals can be used. On the one hand, the F sensing type enables more utilization of WSs. On the contrary, it causes higher delay. Also, conducting F sensing before AC_BE frames maintains the desired priority of these frames.

International Journal of Computer Networks & Communications (IJCNC) Vol.10, No.1, January 2018

Table 2.Sensing strategy based on frame access category

| Access Categories | QoS requirements | Sensing type | Sensing duration ($S_d$) in ms | Sensing characteristics |
|---|---|---|---|---|
| Voice (AC_VO) | Highest priority | Coarse (C) sensing | $S_d \leq 1$ | Cannot distinguish between PU and SU. Poor performance in low SNR. |
| Video (AC_VI) | Second highest priority | Moderate (M) sensing | $1 < S_d \leq 5$ | Cannot distinguish between PU and SU. Moderate sensing accuracy. |
| Best Effort (AC_BE) | Low priority | Fine (F) sensing | $5 < S_d \leq 50$ | Can distinguish between PU and SU. High sensing accuracy. |
| Background (AC_BK) | Lowest priority | Extra Fine (EF) sensing | $S_d > 50$ | Can distinguish between PU and SU. High sensing accuracy even in low SNR. |

As AC_BK frames have the lowest priority, EF sensing should be carried out before them for a duration $S_d$ larger than 50 ms. The EF category allows for the employment of sophisticated sensing methods to achieve high precision, even for the case with low SNR. That is, higher spectrum utilization can be accomplished, but with increased delay.

When F and EF sensing recognize the appearance of PU in the current WS channel, the White-Fi device must scan for other vacant channels and leave the currently occupied one. Otherwise, when the current channel is found busy by other SUs, the device can continue to use and share the current channel with other SU. In the case of C and M sensing, the spectrum assessment outcome cannot be certain about the PU presence. Hence, as long as the appearance for the PU is not certain, other factors should be considered before conducting handoff procedure to another available channel.

## 6. EVALUATIONS OF THE PROPOSED SENSING STRATEGY

In this section, we demonstrate the QoS enhancement that can be achieved by considering the application requirements in selecting the proper sensing strategy. Two scenarios were implemented to compare between fixed sensing approach and selecting the proper sensing based on the QoS requirement approach. The network in both scenarios was the same with four nodes and one server. Three applications, voice, video, and email, were configured to run simultaneously on all nodes for both scenarios. The simulated environment was implemented to reflect a real-life scenario where the wireless device is used to run IP telephony voice application, high-resolution video conferencing and heavy load email application simultaneously. The sensing was conducted for all frames except response frames in both scenarios. The nodes in the first scenario were implemented with the same sensing strategy in Section.0.Hence,fixed sensing duration was used for all different AC frames. In the second scenario, the nodes were implemented to use different $S_d$ based on the AC of the frame to be sent as proposed in Section0.

The first scenario was simulated for four different fixed sensing durations. In each run, the sensing duration was one of these values: 1 ms, 5 ms, 50 ms and 100 ms. As each one of these values fallen in different sensing type proposed in Table .In the second scenario, the nodes were implemented to select $S_d$ based on the AC of the frame, such as $S_d$ = 1 ms for AC_VO, $S_d$ = 5 ms for AC_VI, $S_d$ = 50 ms for AC_BE and $S_d$ = 100 ms for AC_BK. The achieved average throughput for BE frames shown in Figure 6. Demonstrates that the proposed strategy based on selection approach can reach higher throughput compared to fixed sensing approach even when the duration is fixed to 1 ms. The average media access delay, experienced by the voice frames is illustrated in Figure 7. The proposed approach experiences average media access delays that are less than those for the fixed sensing lengths of 50 or 100 ms. The fixed approach with low sensing duration, e.g., 1 ms or 5 ms, results in less medium access delay but with less achieved average throughput compared to the proposed strategy. Moreover, the fixed small sensing duration most likely results in inefficient spectrum utilization and less protection to PU signals that may not comply with regulation requirements. Therefore, the proposed sensing strategy can enhance the achieved QoS of White-Fi devices while maintaining efficient spectrum utilization.

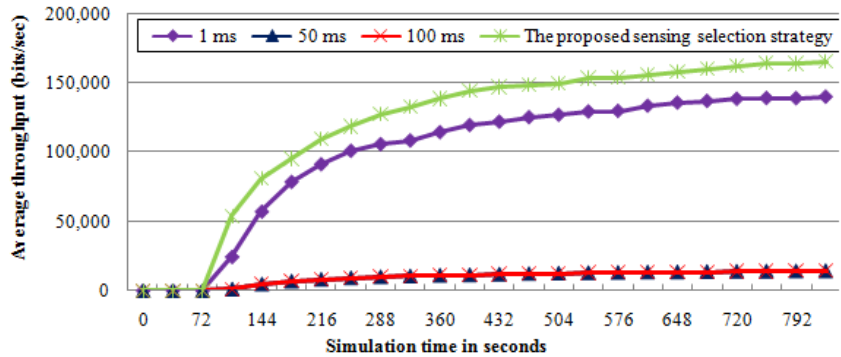

Figure 6. Average throughput for different fixed sensing durations and our proposed sensing strategy (voice, video and email applications running concurrently)

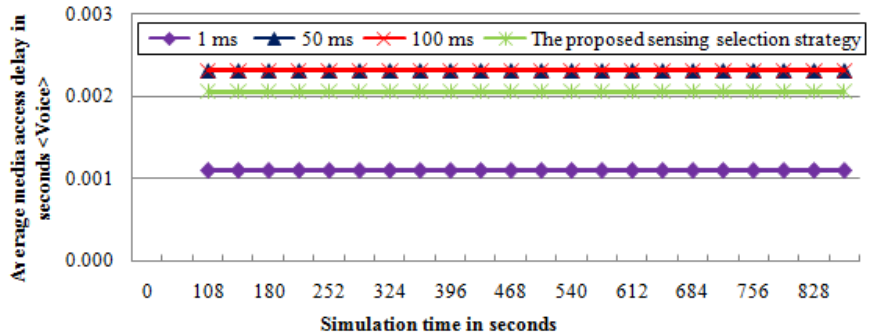

Figure 7. Average voice access delays for different fixed sensing durations and our proposed sensing strategy(voice, video and email applications running concurrently)

## 7. CONCLUSIONS

High detection precision in CR networks may be dependent on increased sensing durations. In this work, we studied the effects of varying sensing durations on the QoS of various applications in the White-Fi networks. Our simulation studies show that compared to video and email traffic, voice signals are more affected by the sensing operations. As such, when the sensing duration is increased to improve detection precision, voice packets experience additional delays, more than the other types of traffic. As a result, the IEE802.11e mechanisms may not be able to achieve the stipulated QoS performance of voice applications requiring small delays. To address this issue, we have proposed a strategy that selects the sensing parameters based on the 802.11e frame categories. The proposed sensing strategy results in QoS improvements while providing high PU protections and efficient spectrum utilization. Our future works aim to develop this sensing strategy further by the inclusion of more parameters, whose effects are combined using artificially intelligent approaches.

## AUTHORS

Nabil Giweli received the B.Sc. degree in Communication Engineering from Tripoli University, Libya, in 1997, the Master degree in Information and Communication Technology (with the dean medal award) from the Western Sydney University in 2011, and another M.Sc. form the same university in Cloud Security in 2013. Currently, he is a Ph.D. candidate and a casual teacher at the School of Computing, Engineering and Mathematics, Western Sydney University, Australia. His current research area is in Cognitive Radio Technologies. 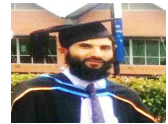

Dr. Seyed Shahrestani completed his PhD degree in Electrical and Information Engineering at the University of Sydney. He joined Western Sydney University (Western) in 1999, where he is currently a Senior Lecturer. He is also the head of the Networking, Security and Cloud Research (NSCR) group at Western. His main teaching and research interests include: computer networking, management and security of networked systems, analysis, control and management of complex systems, artificial intelligence applications, and health ICT. He is also highly active in higher degree research training supervision, with successful results. 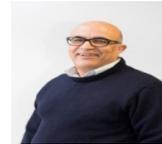

Dr. Hon Cheung graduated from The University of Western Australia in 1984 with First Class Honours in Electrical Engineering. He received his PhD degree from the same university in 1988. He was a lecturer in the Department of Electronic Engineering, Hong Kong Polytechnic from 1988 to 1990. From 1990 to 1999, he was a lecturer in Computer Engineering at Edith Cowan University, Western Australia. He has been a senior lecturer in Computing at Western Sydney University since 2000. Dr Cheung has research experience in a number of areas, including conventional methods in artificial intelligence, fuzzy sets, artificial neural networks, digital signal processing, image processing, network security and forensics, and communications and networking. In the area of teaching, Dr Cheung has experience in development and delivery of a relative large number of subjects in computer science, electrical and electronic engineering, computer engineering and networking. 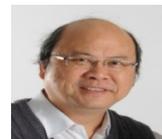